\documentclass[final,5p,times,twocolumn]{elsarticle}

\usepackage{amssymb}
\usepackage{amsmath}

\usepackage{subcaption}
\usepackage{xcolor}

\usepackage{booktabs}
\newcommand{\HOI}{\mathrm{HOI}}

\usepackage{hyperref}

\NewDocumentCommand{\HOIold}{>{\SplitArgument{2}{}}m}{\HOIaux#1}
\NewDocumentCommand{\HOIaux}{mmm}{$\overleftarrow{#1#2}_{#3}$}

\usepackage{duckuments}

\journal{Chaos, Solitons and Fractals}

\begin{document}

\begin{frontmatter}



\title{Modification speed and radius of higher-order interactions\\ alter the oscillatory dynamics in an agent-based model}


\author[KERMIT,BionamiX]{Thomas Van Giel\corref{cor1}}
\cortext[cor1]{Corresponding author}
\ead{thomas.vangiel@ugent.be}

\author[BionamiX]{Hanna Jaspaert}
\author[KERMIT,BionamiX]{Aisling J. Daly}
\author[KERMIT]{Bernard De Baets}
\author[BionamiX]{Jan M.\ Baetens}

\affiliation[KERMIT]{organization={KERMIT, Department of Data Analysis and Mathematical Modelling, Ghent University},
            addressline={Coupure links 653}, 
            city={Gent},
            postcode={9000}, 
            country={Belgium}}

\affiliation[BionamiX]{organization={BionamiX, Department of Data Analysis and Mathematical Modelling, Ghent University},
            addressline={Coupure links 653}, 
            city={Gent},
            postcode={9000}, 
            country={Belgium}}

\begin{abstract}
Understanding the population dynamics of ecological systems is crucial for predicting shifts in biodiversity and ensuring the protection of these systems. Established models often focus on pairwise species interactions, yet recent studies have highlighted the importance of higher-order interactions (HOIs) in shaping community structure and function. In this study, we investigate the effects of HOIs in an agent-based model with three species engaged in intransitive competition. We introduce an HOI where one species modifies the competition between the other two. We explore the impact of the strength, radius of influence, and speed of this interaction modification on species abundances and oscillations thereof. Our results show that these abundances are not only greatly impacted by the strength, but also by the radius and speed of the interaction modification. A deeper investigation demonstrates that the changes in the oscillations are caused by the interaction modification itself, and not the change in pairwise interaction strength caused by the HOI. These results emphasize the importance of considering the spatio-temporal scales of higher-order interactions when assessing ecosystem stability, highlighting that such interactions can introduce complex dynamical behaviors that go beyond the predictions of traditional pairwise or simpler higher-order models.
\end{abstract}

\begin{highlights}
\item Higher-order interactions, which involve more than two species, require significant extensions of ecological modelling theory to tackle these `beyond the pairwise' settings.
\item For the first time, higher-order interactions are considered in an agent-based model, facilitating the investigation of both temporal and spatial effects of such interactions. 
\item The model is set up in such way that the speed and radius of the interaction modification can be independently varied, allowing for a comprehensive analysis of their effects.
\item We adapt Monte-Carlo Singular Spectrum Analysis to compare the oscillatory dynamics of different parameter sets, in order to see the effects of higher-order interactions on ecosystem dynamics.
\item We observe that the speed and radius have profound effects on the oscillation dynamics of species abundances, even if they do not affect the mean species abundances.
\end{highlights}


\end{frontmatter}


\begin{table}
  \caption{List of symbols and abbreviations}
  \label{tab: symbols}
  \begin{tabular}{ll}
    \toprule
    Abbreviation & Description \\
    \midrule
    HOI & Higher-order interaction \\
    ABM & Agent-based model \\
    ODE & Ordinary differential equation \\
    SSA & Singular Spectrum Analysis \\
    EOF & Empirical orthogonal function \\
    MCSSA & Monte Carlo Singular Spectrum Analysis \\
    VMEOF & Proportion of variance explained\\
         & by the main EOF \\
    \midrule
    $\alpha_{ij}$ & Interspecific competition coefficient \\
    $\alpha_{ii}$ & Intraspecific competition coefficient \\
    $P^{\mathrm{c}}_{ij} $ & Probability of competition between agents\\
                  & of different species\\
    $P^{\mathrm{c}}_{ii} $ & Probability of competition between agents \\
                  & of the same species\\
    $\beta$ & Strength of the interaction modification \\
    $\omega$ & Speed of the interaction modification \\
    $R_{\HOI}$ & Radius of the interaction modification \\
    $m$ & Modification value of a specific grid cell \\

    \bottomrule
  \end{tabular}
\end{table}

\section{Introduction}
\label{sec: Intro}
Grasping the dynamics of ecological systems is essential for anticipating changes in biodiversity and developing effective conservation strategies. While traditional models prima\-rily emphasize pairwise species interactions~\cite{lotkaAnalyticalNoteCertain1920,chessonMechanismsMaintenanceSpecies2000,gallienEffectsIntransitiveCompetition2017}, recent research underscores the significant role of higher-order interactions (HOIs) in shaping community structure and ecosystem functioning~\cite{duran-salaStabilityCompetitiveEcological2025,chatterjeeHowCombinedPairwise2024}. An HOI, by definition, is an interaction involving more than two species. Hence, an HOI involving three species can typically be modelled as a pairwise interaction between two species that is modified by the presence of a third species~\cite{kleinhesselinkDetectingInterpretingHigherorder2022,vangielModificationSpeedAlters2025}. This modification can be positive or negative, in the sense of strengthening or weakening the interaction, and of variable magnitude.

Theoretical research on HOIs pursues different avenues, with systems of ordinary differential equations (ODEs) being the most commonly used mathematical models~\cite{gibbsCoexistenceDiverseCommunities2022,chatterjeeControllingSpeciesDensities2022,singhHigherOrderInteractions2021}, whereas data-driven methods have been used to demonstrate the presence of HOIs in real-life systems~\cite{barbosaExperimentalEvidenceHidden2023,bucheMultitrophicHigherOrderInteractions2024}. Recent literature shows that HOIs can have profound effects on species coexistence~\cite{gibbsCoexistenceDiverseCommunities2022,singhHigherOrderInteractions2021} and the general stability of an ecosystem. It has been observed that the introduction of HOIs can lead to coexistence where it would have been impossible without the  modification~\cite{manhartGrowthTradeoffsProduce2018, terryInteractionModificationsLead2019}. Likewise, it has been shown that the introduction of HOIs can either reduce the amplitude of oscillations in species abundances~\cite{duran-salaStabilityCompetitiveEcological2025,shenConnectingHigherorderInteractions2023} or increase them~\cite{vangielModificationSpeedAlters2025, grilliHigherorderInteractionsStabilize2017}, depending on the underlying assumptions.

Although experiments involving HOIs are more complex than those for pairwise interactions, many experiments have been performed to show the presence of HOIs in various different types of real-life systems, such as those involving plants~\cite{kleinhesselinkDetectingInterpretingHigherorder2022, liDirectNeighbourhoodEffects2021}, insects~\cite{barbosaExperimentalEvidenceHidden2023,bucheMultitrophicHigherOrderInteractions2024} and aquatic organisms~\cite{shenConnectingHigherorderInteractions2023}.

Most research on HOIs assumes (typically implicitly) that interactions change as soon as the modifier species is present~\cite{duran-salaStabilityCompetitiveEcological2025,gibbsCoexistenceDiverseCommunities2022, grilliHigherorderInteractionsStabilize2017}. However, in many real-life systems the interaction modification is not instantaneous, but rather takes time to establish itself~\cite{vangielModificationSpeedAlters2025}. Examples are interaction modification through evolution~\cite{patelPartitioningEffectsEcoevolutionary2018,govaertEcoevolutionaryPartitioningMetrics2016}, learning behaviour~\cite{powerWhatCanEcosystems2015} and environmental changes due to ecosystem engineers or invasion~\cite{jonesPositiveNegativeEffects1997}. These examples show that the interaction modification can take time to establish, which means that it does not only depend on the presence of the modifier, but also on the time elapsed since the modifier established itself in the region of the pairwise interaction~\cite{vangielModificationSpeedAlters2025}. This establishment time is an important aspect of HOIs that has received limited attention, and is one of the main focuses of this paper.

Here, we investigate the effects of HOIs on the dynamics of ecological systems using an agent-based model (ABM, also referred to as individual-based model or IBM in other literature). ABMs are powerful tools for simulating many different types of systems, including economic~\cite{baustertUncertaintyAnalysisAgentbased2017,axtellAgentBasedModelingEconomics2025}, social~\cite{celeAgentBasedModeling,alvarez-rodriguezEvolutionaryDynamicsHigherorder2021}, health-related~\cite{pillaiAgentbasedModelingCOVID192024,stuartHPVsimAgentbasedModel2024} and ecological systems~\cite{bampohSimulatingRelativeEffects2021,ferraroEffectsUngulateDensity2022}. They are especially useful for modelling complex systems with many interacting components, such as ecosystems, where individual behaviour can lead to emergent patterns at the population level.

In an ecological setting, two levels of complexity of ABMs are interesting to consider. At one end of the spectrum, there are models that capture all of the necessary complications, interactions and behaviours to model a real-life system accurately enough to draw conclusions applicable to real-life situations. For such models it is important to have a good understanding of the underlying mechanisms driving the system, without overcomplicating them~\cite{sunSimpleComplicatedAgentbased2016, aladwaniEcologicalModelsHigher2020}. At the other end of the spectrum, there are very abstract ABMs, keeping the model as simple as possible in order to better understand the fundamental forces driving emergent behaviour and the underlying mechanisms of the system. These models are usually not applicable to real-life systems, but can be used to advance our understanding of how different mechanisms influence these systems~\cite{sunSimpleComplicatedAgentbased2016, mayUsesAbusesMathematics2004}. In this work, we follow the latter approach to pursue a mechanistic understanding of how HOIs affect the dynamics of a system, using a more individual and spatially explicit way than previous ODE-based research~\cite{gibbsCoexistenceDiverseCommunities2022,chatterjeeControllingSpeciesDensities2022,Letten2019}.

We develop an ABM to simulate a three-species system with intransitive competition and one HOI. First, we describe the model and the different scenarios used in our simulations. Then we examine the effect of the different parameters on the mean abundances of the three species, and how these parameters affect oscillations of species abundances in the system. Such oscillations are important for the stability of an ecosystem, as they can indicate greater vulnerability to perturbations and a higher risk of extinction~\cite{heinoSynchronousDynamicsRates1997, myersPopulationCyclesGeneralities2018}. However, oscillations in species abundances have also been shown to have positive effects, such as increased species coexistence~\cite{petterssonSpatialHeterogeneityEnhance2021}, by shifting resource use~\cite{shimadzuDiversityMaintainedSeasonal2013} or stabilising ecosystem functions~\cite{waggDiversityAsynchronySoil2021, wilcoxAsynchronyLocalCommunities2017}. Finally, we study how the interaction modification itself affects these oscillations, and how this differs from simply changing the pairwise interaction strength. This will allow us to disentangle the effects of the HOI from those of the pairwise interactions. 

\section{The agent-based model}

The model used in this paper is an ABM where each agent is assigned to one of three species~$A$, $B$ and $C$. Agents move randomly around the grid, interact with each other and randomly reproduce. A full ODD description~\cite{grimmODDProtocolReview2010} of the model can be found in the supplementary information (Sec.~S1); a short summary is given here. 

\subsection{Competition structure}
\begin{figure}[ht!]
  \centering
  \includegraphics[width=0.2\textwidth]{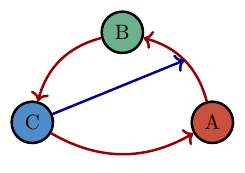}
  \caption{Graph of the competition model. The circles indicate the three species~$A$, $B$ and $C$. The red arrows indicate the pairwise competition interactions. The blue arrow between species~$C$ and the interaction between species~$A$ and $B$ indicates that species~$C$ modifies the competition between species~$A$ and $B$.}
  \label{fig: intransitive HOI graph}
\end{figure}

The pairwise competition graph of the three species is intransitive, as shown by the red arrows in Fig.~\ref{fig: intransitive HOI graph}, indicating that species~$A$ is superior to species~$B$, species~$B$ to species~$C$ and species~$C$ to species~$A$. A single HOI is also present, which is modelled as an interaction modification, with species~$C$ modifying the competition between species~$A$ and~$B$, as indicated by the blue arrow in Fig.~\ref{fig: intransitive HOI graph}. For this reason, we refer to species~$C$ as the \emph{modifier}.

\subsection{Agent behaviour}
\label{sec: agent behaviour}

Simulations take place on a 200 x 200 grid, where multiple agents of the three species are present. At most one agent can occupy a given grid cell. The agents move around according to a random walk and can reproduce randomly at each time step with probability $r = 0.005$. When two agents move to the same grid cell at a given time step, they compete with a given probability $P^{\mathrm{c}}$ (Sec.~\ref{sec: probability of competition}). However, if the interaction between the two agents is modified by the presence of the modifier species~$C$, then the probability $P^{\mathrm{c}}$ is adjusted (Sec.~\ref{sec: interaction modification}).  If competition occurs, then the winner survives, whereas the loser is removed from the grid.

\subsection{Pairwise competition}
\label{sec: probability of competition}
If two agents attempt to move to the same grid cell, they either compete with probability $P^{\mathrm{c}}$, causing the loser of the competition to die and be removed from the grid, or they swap places with probability $1 - P^{\mathrm{c}}$. If the competition occurs between two agents of the same species, the probability of competition is given by
\begin{equation}
  P^{\mathrm{c}}_{ii}  = \tanh \left(\alpha_{ii}\right)\,,
\end{equation}
where $\alpha_{ii} > 0$ is the intraspecific competition coefficient of species $i$, a constant parameter in the model, identical for all species. The winner of the competition is chosen arbitrarily. The hyperbolic tangent function ensures a value for the probability function between~$0$ and~$1$. 

If the interaction takes place between two agents of different species and is not modified by the presence of the modifier (\emph{e.g.,} between agents of species~$A$ and $C$), then the probability of competition is given by
\begin{equation}
  P^{\mathrm{c}}_{ij} = \tanh \left(\alpha_{ij} \right)\,,
\end{equation}
where $\alpha_{ij} > 0$ is the pairwise competition coefficient between species~$i$ and $j$, with $i \neq j$.
The winner of the competition event is determined by the intransitive competition graph (Fig.~\ref{fig: intransitive HOI graph}).

\subsection{Interaction modification in the ABM}
\label{sec: interaction modification}
\begin{figure}[ht!]
  \centering
  \begin{subfigure}{.15\textwidth}
    \centering
    \includegraphics[width = \textwidth]{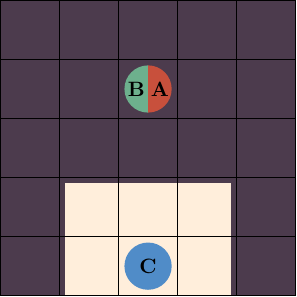} 
    \caption{}
    \label{fig: ABM no modification}
  \end{subfigure}
  \begin{subfigure}{.15\textwidth}
  \centering
  \includegraphics[width = \textwidth]{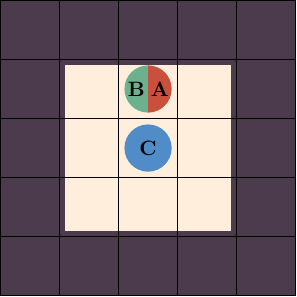}
  \caption{}
  \label{fig: interaction modification example}
\end{subfigure}
\caption{Any interaction between red ($A$) and green ($B$) agents is modified if it occurs in the neighbourhood of a blue ($C$) agent. The half-green, half-red circle represents the grid cell where agents of species~$B$ and $A$ meet and interact. The blue circle represents an agent of the modifier species~$C$. (a) No modification happens, since the interaction occurs outside of radius of modification of the agent of species~$C$. (b) The interaction is modified when it occurs within the radius of modification of an agent of species~$C$.}
\end{figure}

Species~$C$ modifies the pairwise competition interactions between species~$A$ and $B$. If an agent of species~$A$ and an agent of species~$B$ interact within the radius of modification $R_{\HOI}$ of an agent of species~$C$, then the pairwise interaction will be modified by changing the probability of competition $P^{\mathrm{c}}_{AB}$ between the two species (Fig.~\ref{fig: interaction modification example}). 

The interaction modification is defined by two parameters: the strength ($\beta$) and the radius ($R_{\HOI}$) of the modification. The strength of the modification $\beta$ determines to what extent the probability of competition is modified in the neighbourhood of the modifier. It can be positive or negative, ranging from $-\infty$ to $+\infty$. The radius of the modification $R_{\HOI}$ determines the distance from the modifier within which pairwise interactions are modified. We use the Chebyshev distance (or $L_\infty$-metric) to define the neighbourhood of the modifier. This means that the neighbourhood of the modifier is a square with side length $2 R_{\HOI} + 1$ cells. The radius $R_{\HOI}$ can range from 1 to 100, where at the upper limit the entire grid becomes the modification area.

\begin{figure}[ht!]
  \centering
  \includegraphics[width=0.45\textwidth]{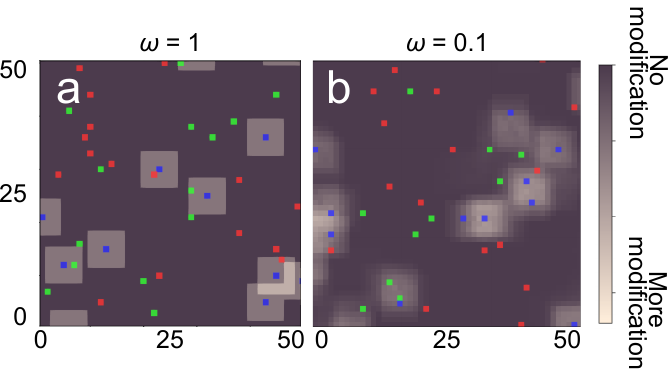}
  \caption{A visualisation of the modification caused by species~$C$ in the ABM with $R_{\HOI} = 3$. The three species~$A$, $B$, and $C$ are shown in red, green and blue, respectively. Lighter regions are the cells at which an interaction between $A$ and $B$ is modified. The lighter the grid cell, the more intense the modification. Only a 50 x 50 subgrid of the entire 200 x 200 grid is shown here for clarity.}
  \label{fig: typical simulation}
\end{figure}

Recent research has shown that the speed at which the HOI manifests itself is also an important factor in determining the behaviour, specifically the oscillatory dynamics, of the system~\cite{vangielModificationSpeedAlters2025}.  We can incorporate this temporal aspect into our ABM by adding a third parameter, namely the speed of the modification ($\omega$). This parameter determines how fast the modification of the interaction between two agents is established. Specifically, when $\omega$ is high ($\omega = 1$), the modification always occurs at its full strength, around the position of the agents of species~$C$ at the end of the time step. If all agents of species~$C$ disappear, the modification disappears as well. If $\omega$ is lower ($ 0 \leq \omega < 1$), then the modification takes some time to reach its full strength in the presence of species~$C$, and it also disappears more gradually from an area where agents of type~$C$ have disappeared. Thus, the modification can have a delayed effect (Fig.~\ref{fig: typical simulation}).

We can model the effects of the interaction modification by changing the probability of competition when 
agents of species~$A$ and $B$ meet, by incorporating the (historical) presence of the modifier species~$C$:
\begin{equation}
  \label{eq: probability of competition modified}
  P^{\mathrm{c}}_{ij}  = \tanh \left(\alpha_{ij}  m_t^{p} \right)\,,
\end{equation}
where $m_t^{p}$ is the modifier strength at the cell at position $p$ at which the agents meet at time $t$. The modifier strength $m_t^p$ changes the probability of competition $P^{\mathrm{c}}$ between species~$A$ and $B$ if the competition occurs at the focal cell at position~$p$. This value depends on the (historical) presence of the modifier and the HOI parameters $R_{\HOI}$ and $\beta$. The modification value at time step $t$ for a grid cell at position $p$ ($m_{t}^p$) is updated according to:
\begin{equation}
  \label{eq: m update}
  m_{t}^{p}  = m_{t-1}^{p}  + \omega \left(1 - m_{t-1}^{p}  + \beta^* n_{C}\right) \,,
\end{equation}
where $n_{C}$ is the number of agents of species~$C$ in the neighbourhood of the focal cell. Here, $\beta^*$ is a rescaled version of $\beta$:
\begin{equation}
  \label{eq: beta star}
  \beta^* = \beta\frac{9}{\left( 2 R_{\HOI} + 1\right)^2},
\end{equation}
where the denominator is the number of cells in the neighbourhood of the focal cell within radius $R_{\HOI}$ and the numerator is the number of cells in the neighbourhood within radius $R_{\HOI} = 1$. The use of $\beta^*$ ensures that the total amount of modification caused by a modifier is independent of $R_{\HOI}$.  In this way, we try to ensure that $R_{\HOI}$ has no effect on the average interaction strength between individuals, and thus that $R_{\HOI}$ has no effect on the mean abundances of the species. All $m_t$ are initialised at $m_0 = 1$, which is the value at which no modification occurs. Equation~\eqref{eq: m update} is the discretised version of negative exponential decay, often used to model the decay of natural phenomena~\cite{cornwellDecompositionTrajectoriesDiverse2014, olsonEnergyStorageBalance1963}. 

The value of any $m_t^p$ can be negative, depending on the value of~$\beta$ and the number of agents of species~$C$ in the neighbourhood of the grid cell. This means that~$P^{\mathrm{c}}$ can also become negative, which would cause the outcome of the competition to be reversed, and the agent of species~$B$ would defeat the agent of species~$A$ with probability~ $|P^{\mathrm{c}}|$.

Equation~\eqref{eq: m update} is a discretised version of the relationship proposed in~\cite{vangielModificationSpeedAlters2025}, where it was shown that $\omega$ has no effect on the mean abundances of the species, but did have an effect on the formation of oscillations in the system. In this study as well, the steady state value of the modification value of a grid cell~$m_t^p$ is independent of $\omega$, and thus we expect that the mean abundance of the species will be unaffected by~$\omega$. It can be easily proven that, in the absence of agents of species~$C$, for $\omega \leq 1$, the modification values $m_t^p$ will converge to~1 (the value at which no modification occurs) as time goes to infinity (see supplementary Section~S3). The speed with which the values of~$m_t^p$ converge is determined by $\omega$.

For instantaneous modification ($\omega = 1$), Eq.~\eqref{eq: m update} simplifies to:
\begin{equation}
  m_{t+1}^p = 1 + \beta^* n_{C}\,,
\end{equation}
where the modification value is directly proportional to the number of agents of species~$C$ in the neighbourhood (determined by $R_{\HOI}$) of the cell at  time~$t$ and position $p$, irrespective of its previous values $m_t^p, m_{t-1}^p, \ldots $. In the case of a slower modification ($\omega < 1$), the modification value at time $t$ also depends on $m_t^p, m_{t-1}^p, \ldots $. This means that the modification value also depends on the number of agents of species~$C$ in the neighbourhood of the cell at previous time steps. It thus takes longer for the modification to establish itself and for it to disappear once agents of species~$C$ are no longer present.

\subsection{Default parameter values and number of replicates}
In the remainder of this paper, we will use the following parameter values when referring to the base HOI scenario: 
\begin{table}[h!]
\centering
\caption{The parameter values of the base scenario.}
\label{tab: base parameters}
\begin{tabular}{|l|c|c|c|c|c|}
    \hline
    Parameter & $\alpha_{ij}$ & $\alpha_{ii}$ & $\beta$ & $R_{\mathrm{HOI}}$ \\ \hline
    Value     & 0.3 & 0.4 & $ 0$ & 3 \\ \hline
\end{tabular}
\end{table}

These parameter values lead to a pairwise system with oscillating abundances and consistent coexistence of the three species. The simulations are all initiated with 200 agents of each species, randomly distributed over the 200 x 200 grid. The agents then move around the grid for 10~000 time steps before the data collection starts, to avoid transient effects due to the initial conditions (Sec.~S6). After this initialisation phase, the model is run for 30~000 time steps which ensures that the system has reached a quasi-stationary state and has run sufficiently long to capture any oscillatory behaviour.

A consistency analysis of the system was carried out according to~\cite{hamisUncertaintySensitivityAnalyses2020} in order to determine how many model runs are necessary to mitigate uncertainty originating from the model stochasticity (Sec.~S2). This analysis showed that 100 repetitions are sufficient to capture the uncertainty due to stochasticity in the data.


\subsection{Simulation procedure}
A simulation starts with initialising the 200 x 200 grid with 200 agents of each species, randomly distributed over the grid. 
Then at each time step $t$:
\begin{enumerate}
  \item A random agent is chosen from the grid.
  \item A random grid cell within radius 1 of the current location of the agent is selected.
  \item If the grid cell is unoccupied, then the agent reproduces (a new agent of the same species is created) at that grid cell with probability $r = 0.005$, or the agent moves to that grid cell with probability $1-r$.
  \item If the grid cell is occupied by another agent, then the two agents compete with probability $P^{\mathrm{c}}$ (Sec.~\ref{sec: probability of competition}) or swap  places with probability $1-P^{\mathrm{c}}$. If the interacting agents are from species~$A$ and $B$, then $P^{\mathrm{c}}$ is determined by the modification value of the cell $m_t^p$ (Sec.~\ref{sec: interaction modification}).
  \item Choose another agent at random from the grid and repeat Steps 2--4 until all agents have moved.
  \item The values of $m_t^p$ are updated according to Eq.~\eqref{eq: m update} for all grid cells.
\end{enumerate}

 The simulation then proceeds to the next time step $t + 1$ and at every time step, the abundances of the three different species are tracked, as well as the modification value of every grid cell. The simulation stops after 30~000 time steps.

\section{Results and discussion}
\subsection{Species abundances}
The first step in our analysis concerns the effect of the model parameters on the population dynamics. Specifically, we look at the effect of the HOI on the mean abundance of the three species. For this purpose, a Sobol analysis~\cite{sobolGlobalSensitivityIndices2001} is performed. Sobol analysis is a variance-based sensitivity analysis method that decomposes output variance into contributions from individual input parameters, allowing the relative importance of each parameter to be quantified. Here, we focus on the species population means.
Since we want to examine the effect of the HOI on the model dynamics, the input parameters are those related to the interaction modification, being $\beta$, $\omega$ and $R_{\HOI}$. 

\begin{figure}[ht!]
  \centering
  \includegraphics[width=0.45\textwidth]{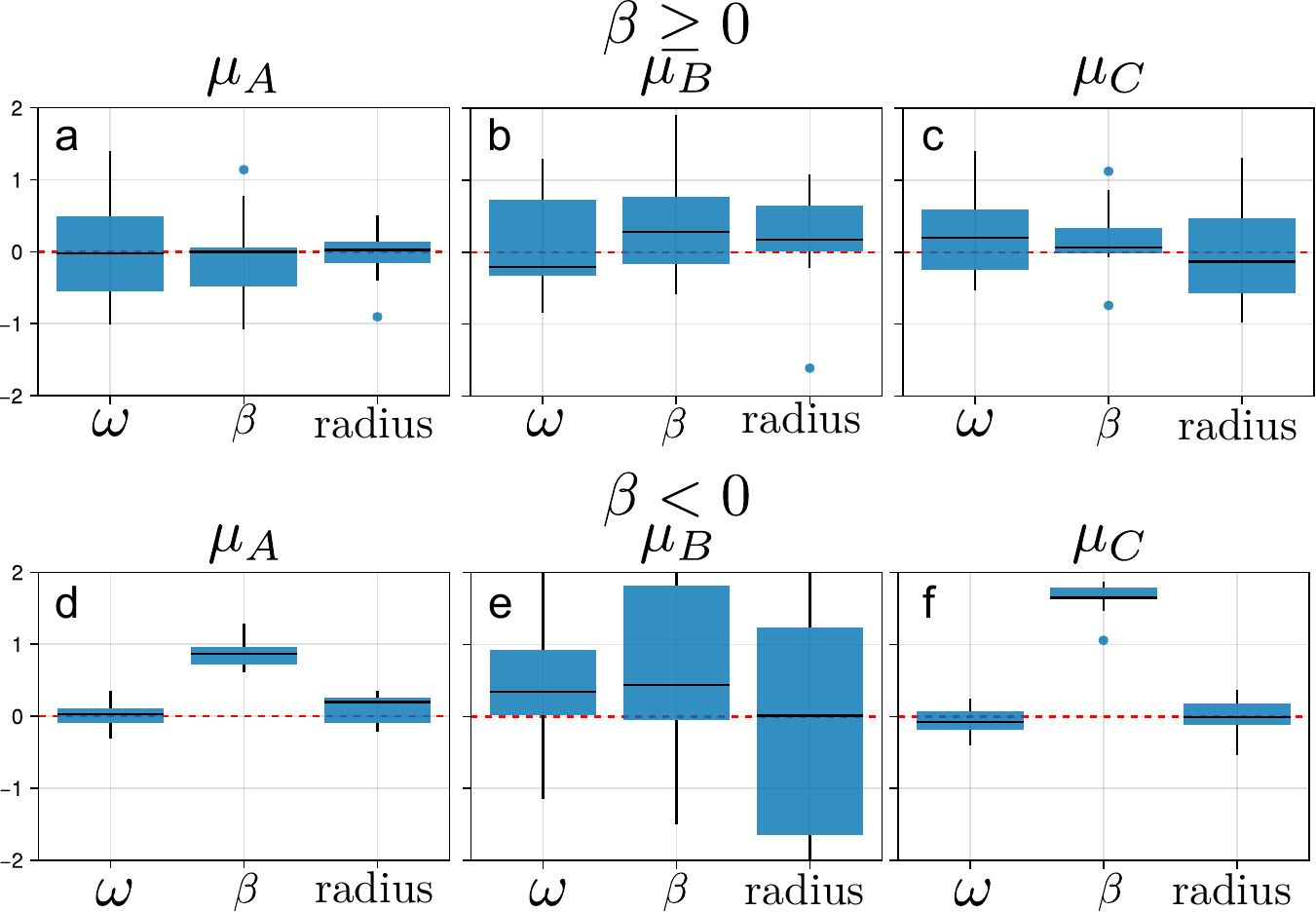}
  \caption{Boxplots of the effect of the HOI strength ($\beta$), modification speed ($\omega$) and modification radius ($R_{\HOI}$) on the mean abundance of species~$A$ ((a) and (d)), species~$B$ ((b) and (e)) and species~$C$ ((c) and (f)), for positive modification ((a)-(c)) and negative modification ((d)-(f)).}
  \label{fig: SOBOL results}
\end{figure}

Since positive (strengthening) and negative (weakening or reversing) values of $\beta$ are very different types of modification, we construct two separate Sobol sequences on the hypercube of parameter ranges: one for $\beta \in [-25, 0]$ and another one for $\beta \in [0, 25]$, while keeping $\omega \in [0.001, 1]$ and $R_{\HOI} \in [1, 100]$ in both cases. This ensures that the samples are evenly distributed over the respective parameter spaces. For each sequence, the model is run at every sampled parameter set, and the mean abundances of species~$A$, $B$ and $C$ ($\mu_A$, $\mu_B$ and~$\mu_C$) are computed. This process is repeated 100 times to account for model stochasticity (Sec.~S4). The results of this analysis are presented in Figure~\ref{fig: SOBOL results}.

In this figure we see that $R_{\HOI}$ and $\omega$ do not have a notable effect on the means of any of the abundances, independent of $\beta$. This validates the model formulation, since the model was built for $\omega$ and $R_{\HOI}$ to have no effect on the mean abundances of the populations (Sec.~\ref{sec: interaction modification}). 

For a positive modification ($\beta > 0$), we can see that $\beta$ does not affect the mean abundances of any of the three species. However, for a negative modification ($\beta < 0$), we can see that $\beta$ has a pronounced effect on both~$\mu_{A}$ and~$\mu_{C}$. This is because the modification changes the probability of competition between two agents of species~$A$ and~$B$. With positive $\beta$, the modification strength~$m$ of a cell will always be positive. With positive $m$, the probability of competition when an agent of species~$A$ and species~$B$ meet is increased. When $\beta$ is slightly negative, $m$ will typically take a value between zero and one, and thus the probability of competition decreases. When $\beta$ is sufficiently negative, $m$ can become negative and the competition changes direction. This means that, with a modified interaction, an agent of species~$B$ would defeat an agent of species~$A$. This change in interaction outcome is likely the driver of the significant change in the mean abundances of species~$A$ and $C$.

\begin{figure}[h!]
  \centering
  \includegraphics[width=0.45\textwidth]{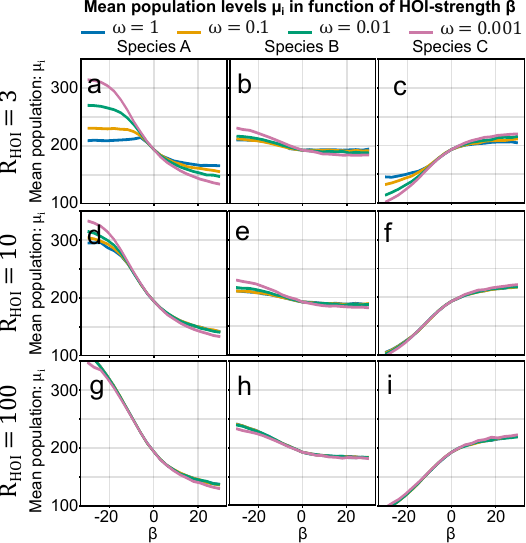}
  \caption{The effect of the modification strength $\beta$ on the mean abundances of species~$A$, $B$ and $C$. The (very small) error bars represent the 95\% confidence interval of the mean population abundance. The results are shown for (a)-(c) $R_{\HOI} = 3$, (d)-(f) $R_{\HOI} = 10$ and (g)-(i) $R_{\HOI} = 100$, for species~$A$, $B$ and $C$ in the first, second and third column. The results are shown for $\omega$ in $\{1, 0.1, 0.01, 0.001\}$.}
  \label{fig: effect beta abundance}
\end{figure}

Because the Sobol analysis does not provide a full picture, we plot the average over 100 simulations of the mean abundances per species, as a function of $\beta$ (Fig.~\ref{fig: effect beta abundance}).
An initial observation is that the error bars are extremely small. This is because the mean abundances are the mean over 100 repetitions, with 30~000 time steps per repetition.

Figure~\ref{fig: effect beta abundance} provides a clearer picture of the combined effects of the different parameters, especially the combination of $\omega$ and~$R_{\HOI}$. Firstly, we can see that the mean abundances of all three species are largely independent of $\omega$ and $R_{\HOI}$ for most combinations of these parameters. Only for $\beta < 0$ and $R_{\HOI} = 3$, we can see that $\omega$ has an effect on~$\mu_A$, and to a lesser extent on~$\mu_C$. We can see that if the value of $\omega$ becomes lower, the behaviour of the system becomes more similar to the behaviour observed for higher values of $R_{\HOI}$, and when $\omega$ is higher, the effect of the HOI is reduced.

Small values of $\omega$ cause the modification to change gradually; as modifier individuals move, the resulting modification becomes more spatially dispersed. This is similar to the case of a high $R_{\HOI}$, where the modification value is dispersed over a larger area. For high values of $\omega$ and small values of $R_{\HOI}$, the modification is much more concentrated around the agents of species~$C$, which reduces the chance that a modified cell is the location for an interaction, and thus reduces the effect of the modification on species~$A$ and $B$. In Fig.~\ref{fig: effect beta abundance}a and \ref{fig: effect beta abundance}c, we can see that the effect of the interaction modification does not increase much for very positive or very negative values of~$\beta$. Looking at Eqs.~\eqref{eq: probability of competition modified} and~\eqref{eq: m update}, we can see that $P^{\mathrm{c}}$ converges to $1$ quickly for high values of $\omega$ and low values of $R_{\HOI}$. For very positive or very negative values of~$\beta$, $\vert \alpha_{ij} m_t^p \vert$ will be large, and $P^{\mathrm{c}}_{AB} $ will quickly approach $1$. If $\vert \beta \vert$ increases even further, then $P^{\mathrm{c}}_{AB}$ will not do so, and thus the effect of the modification on the mean abundances will not change much either.

Two more interesting observations can be made from Fig.~\ref{fig: effect beta abundance}. First, for negative values of~$\beta$, the mean abundance of species~$A$ increases with decreasing values of~$\beta$, whereas the mean abundance of species~$C$ decreases. The mean abundance of species~$B$ increases only slightly with decreasing $\beta$. This is somewhat counter-intuitive, since a negative value of~$\beta$ means that the probability of species~$A$ winning against species~$B$ is decreased, and if $\beta$ is sufficiently negative, the probability of species~$B$ winning against species~$A$ is increased. Thus, one would expect that the mean abundance of species~$A$ would decrease and that of species~$B$ would increase with decreasing values of~$\beta$. However, the opposite is true for species~$A$, while species~$B$ stays relatively constant compared to species~$A$ and $C$. A possible explanation for this behaviour is as follows. With decreasing values of $\beta$, the interaction strength between species~$A$ and $B$ decreases, which increases the abundance of species~$B$. This in turn negatively affects species~$C$, whose abundance decreases. The decreased abundance of species~$C$ reduces the competitive pressure on species~$A$, for which the abundance increases. This increases the competitive pressure on~$B$, which counterbalances the original reduction in pressure.

Second, the mean abundance of species~$C$ stays relatively constant for positive values of~$\beta$, independent of~$\omega$ and~$R_{\HOI}$. This is consistent with previous research using models based on differential equations~\cite{vangielModificationSpeedAlters2025}, where it was shown that, for the case where the modification only strengthens or weakens the negative effect of species~$A$ on species~$B$ (there written as \HOIold{BAC}), the mean abundance of species~$C$ remains constant as a function of~$\beta$. In our ABM, when $\beta > 0$, only the probability of~$B$ dying upon encountering $A$ is modified, so we have a similar situation as \HOIold{BAC}. 
However, in~\cite{vangielModificationSpeedAlters2025}, for \HOIold{BAC}, it was shown that the mean abundances of $A$ and $B$ are equal to each other, which is clearly not the case in our agent-based approach. More precisely, in~\cite{vangielModificationSpeedAlters2025}, the population of a competitively superior species grows in the presence of a competitively inferior species (\emph{e.g.,} the abundance of species~$A$ grows in the presence of species~$B$), while in our ABM, the abundance of the superior species~$A$ is not positively impacted by the presence of~$B$, leading to different equilibria.

\subsection{Oscillations in the population abundances}
Even though~$\omega$ has very little effect on the mean population abundances, previous research has shown that it can have an effect on the oscillatory dynamics of the population abundances~\cite{vangielModificationSpeedAlters2025}, and we anticipate that the same could hold for $R_{\HOI}$. To analyse the oscillatory dynamics of our system, a Monte Carlo Singular Spectrum Analysis (MC-SSA) was performed. This is a method that distinguishes oscillations from random noise through a statistical test based on surrogate data generated by Monte Carlo sampling~\cite{allenMonteCarloSSA1996}. It has been used in time series analysis concerning climate~\cite{loeuilleIntrinsicClimaticFactors2004} and population dynamics~\cite{colonBifurcationAnalysisAgentbased2015}. We expect that, although $\omega$ and $R_{\HOI}$ have little effect on the mean abundances of the species, they could have an effect on the oscillations of the abundances in the system.

\subsubsection{Monte Carlo Singular Spectrum Analysis}
In SSA, we first generate the $M \times M$ lag-covariance matrix~$C_X$ (Sec.~S4.1) of a time series $\{X(t)\mid t=1,2, \ldots, N\}$, after which we can diagonalise $C_X$: 
\begin{equation}
C_{X} =E_{X}^T \Lambda_{X} E_{X}\,,
\end{equation}
where $\Lambda_X = \text{diag}(\lambda_1, \lambda_2, \ldots, \lambda_M)$ with $\lambda_i > \lambda_{i+1} > 0$, for $i \in {1, \ldots ,M-1}$, are the real eigenvalues of $C_X$, $E_X \in \mathbb{R}^{M \times M}$ is a matrix of which  the $k$-th column $\rho_k$ is the  eigenvector corresponding to $\lambda_k$.
The different eigenvectors are also called empirical orthogonal functions (EOFs), and they represent different oscillating functions with different frequencies. Projecting the time series $X(t)$ onto each EOF yields the corresponding principle component~\cite{ghilAdvancedSpectralMethods2002}.
 The eigenvalue $\lambda_k$ gives the fraction of the total variance of the time series that is attributed to the $k$-th EOF. Thus, the higher $\lambda_1$ and $\lambda_2$, the more dominant the frequency of the corresponding EOFs. MC-SSA repeatedly executes this method in order to obtain confidence intervals for the magnitude of the \emph{proportion of the variance captured by the main EOFs} (VMEOF). This method is typically used to distinguish oscillations from random noise in a time series. If the VMEOF of the time series is significantly higher than that of randomly generated red noise (Sec.~S4.2), it can be assumed that there are structural oscillations underlying the data, independent of stochasticity. Here, we use this method to compare the strength of oscillations among systems with different parameters. The higher the VMEOF of the time series, the higher the amplitude of the oscillations in the system.

\subsubsection{Effect of the modification strength and radius}
\begin{figure}[ht!]
  \centering
  \includegraphics[width=0.45\textwidth]{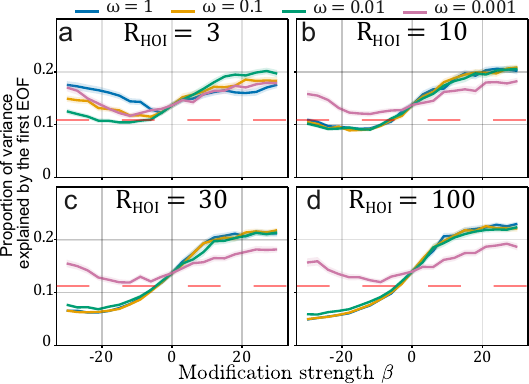}
  \caption{Effect of $\beta$ on the oscillatory behaviour of the abundance of species~$A$, explained by the main EOF, for (a) $R_{\HOI} = 3$, (b)  $R_{\HOI} = 10$, (c) $R_{\HOI} = 30$, (d) $R_{\HOI} = 100$ and for $\omega\in \{
  0.001, 0.01, 0.1, 1\}$. The horizontal, red, dashed lines denote the 95\% confidence intervals of the mean of the proportion of variance explained by the main EOF (VMEOF) as calculated using MC-SSA. The red dotted line denotes the VMEOF of red noise: if the VMEOF of the system is significantly higher, there is a strong statistical difference in the oscillations present in the system compared to random noise.}
  \label{fig: beta MCSSA}
\end{figure}

The effect of the modification strength $\beta$ on the oscillations of the system was analysed for different values of $\omega$ and $R_{\HOI}$. These effects on the VMEOF of the systems are shown in Fig.~\ref{fig: beta MCSSA}.
 
An initial observation is that, for very slow modifications ($\omega = 0.001$), the VMEOF of the time series (thus the magnitude of the oscillations) shows the same behaviour for all values of~$R_{\HOI}$. This is because for very small $\omega$, a modifier species that moves through a given area of the grid leaves behind a trail of modification potential. This causes the area of modification potential to be more scattered in space. For values of $\omega \approx 0$, the modification is homogeneous over the entire grid and is very stable over time, so the VMEOF does not vary with $R_{\HOI}$. 

We can also see that for all combinations of $\omega$ and $R_{\HOI}$, and $\beta > 0$, the VMEOF rises with increasing $\beta$. This is because the modification strength $m$ increases with increasing $\beta$ (Eq.~\eqref{eq: m update}), which causes the probability of competition between species~$A$ and $B$ to increase. This implies that the interaction between species~$A$ and $B$ becomes more pronounced, which implies growing amplitudes of the population oscillations. 

For negative values of $\beta$, the results are more complex. For $R_{\HOI} = 3$ and to a lesser extent for $R_{\HOI} = 10$, Fig.~\ref{fig: beta MCSSA} shows that the VMEOF first decreases with decreasing $\beta$, but then increases again for very negative values of $\beta$. Part of this behaviour can be explained by the fact that, for slightly negative values of $\beta$ (if $0 < m_t^p < 1$ in Eq.~\eqref{eq: m update}, which occurs in equilibrium for $-5.44 < \beta < 0$ when $R_{\HOI} = 3$), the modification strength $m$ is still positive, and thus the probability of competition between species~$A$ and $B$ is decreased. This causes the structural oscillations to become weaker. However, for very negative values of $\beta$, the modification strength $m$ can become negative, which causes competition between species~$A$ and $B$ to be reversed in the presence of~$C$. When $\beta$ is sufficiently negative, decreasing it further increases the probability of competition, which in turn increases the VMEOF.

For higher values of $R_{\HOI}$, the results are different. For $R_{\HOI} = 100$, the VMEOF decreases with decreasing $\beta$, since the modification is dispersed more over the neighbourhood of the modifier as $R_{\HOI}$ grows. This means that the effect of the modifier on the probability of competition between species~$A$ and $B$ is less pronounced, and thus the oscillations weaken.

A very interesting observation for $R_{\HOI} > 10$ is that the VMEOF of the time series for $\omega = 0.001$ does not change much as a function of $\beta$, compared to higher values of $\omega$. As mentioned before, for $\omega = 0.001$, the modification strength $m$ of every cell is very stable over time. Thus the system dynamics align more with those of a system without HOI, as the pairwise interactions stay relatively constant over time. Hence, the VMEOF of the time series with $R_{\HOI} > 10$ and $\omega = 0.001$ is relatively similar for all values of $\beta$ to the VMEOF of the time series without HOI ($\beta = 0$). For the other values of $\omega$, the increase or decrease in oscillation strength due to $\beta$ is very noticeable. Although the average steady state amount of modification in each cell is the same for all values of $\omega$, the effects are much more pronounced for higher values of $\omega$. This leads us to conclude that the oscillations or the dampening thereof are specifically caused by the HOI and to a lesser extent by the change in interaction strength on a pairwise level.

\begin{figure}[ht!]
  \centering
  \includegraphics[width=0.45\textwidth]{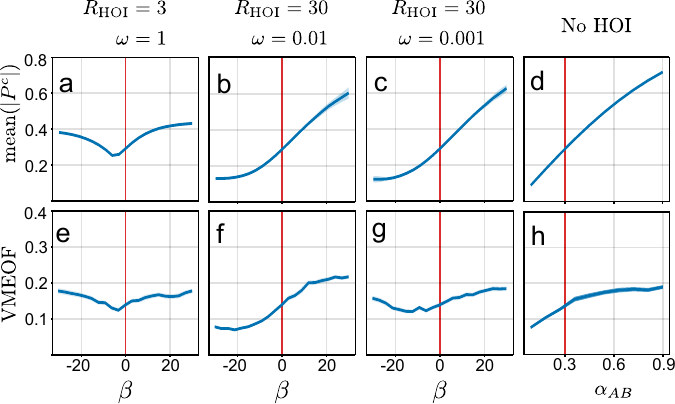}
  \caption{A comparison between the average absolute value of the modification strength over all grid cells as a function of $\beta$ (top row), and the VMEOF of the abundances time series (bottom row). (a) and (e): a system with parameters $R_{\HOI} = 3$ and $\omega = 1$, (b) and (f) $R_{\HOI} = 30$ and $\omega = 0.01$, (c) and (g) $R_{\HOI} = 30$ and $\omega = 0.001$, respectively. (d) and (h) show the results for a system without an HOI present, but where the value of the pairwise interaction strength between species $A$ and $B$ ($\alpha_{AB}$) is changed. The vertical red lines in the figures show the parameters of the baseline scenario (Tab.~\ref{tab: base parameters}). All systems have the same parameters on the red line.}
  
  \label{fig: avg modification strength}
\end{figure}

This is further confirmed by plotting the average modification strength over all grid cells, as a function of $\beta$ (Fig.~\ref{fig: avg modification strength}).

This figure shows that for cases where $\omega$ is not very low, as illustrated by the representative results in Figs.~\ref{fig: avg modification strength}a, b, e and f, the shape of the average modification strength and the VMEOF follow very similar patterns as a function of $\beta$. However, for the case where $\omega = 0.001$, the shape of the average modification strength as a function of $\beta$ is very comparable to the shape where $\omega = 0.01$. However, the shape of the VMEOF as a function of $\beta$ is very different. Since, for very low values of $\omega$, the modification strength $m$ becomes more stable over time, the system behaves more like a system without HOI, this once again shows that the oscillations or the dampening thereof are specifically caused by the HOI, and not by the change in strength of the pairwise interactions. We can also see this in Figs.~\ref{fig: avg modification strength}d and h, which show the interaction strength between $A$ and $B$ and the VMEOF of the abundance time series for a system without HOI. In these figures, instead of adding an HOI, we change the pairwise interaction strength $\alpha_{AB}$ between species $A$ and $B$, thus changing the probability of competition $P^{\mathrm{c}}$ between those two species. We can see that the pairwise case without HOIs and the case with very low values of $\omega$ are very similar. While the average chance of competition between $A$ and $B$ increases, the VMEOF does not increase to the same level as in the system with $R_{\HOI} = 30$ and $\omega = 0.01$, once again showing that the HOI itself is also responsible for oscillations, not just the change in pairwise interaction strength caused by the HOI. For $\alpha_{AB} < 0$, species $B$ wins the competition with $A$, causing its population to increase too much, which leads to the extinction of species $C$ and $A$.

\subsubsection{Effect of the modification speed \texorpdfstring{$\omega$}{omega}}

\begin{figure}[ht!]
  \centering
  \includegraphics[width=0.45\textwidth]{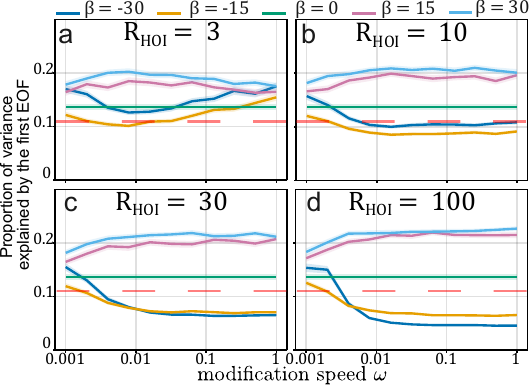}
  \caption{Effect of $\omega$ on the oscillatory behaviour explained by the main EOF, for (a) $R_{\HOI} = 3$, (b)  $R_{\HOI} = 10$, (c) $R_{\HOI} = 30$, (d) $R_{\HOI} = 100$ and for $\beta\in\{-25, -15, -5, 0, 5, 15,25\}$. The dotted lines denote the 95\% confidence intervals of the mean of the proportion of variance explained by the main EOF (VMEOF). The red dotted line denotes the VMEOF of red noise.}
  \label{fig: omega MCSSA}
\end{figure}

To examine the link between the modification speed $\omega$ and the oscillatory dynamics of the abundances, we compare the VMEOF of the time series as a function of $\omega$, for different values of $\beta$ and $R_{\HOI}$ (Fig.~\ref{fig: omega MCSSA}).

Similarly to Fig.~\ref{fig: beta MCSSA}, we see that the curves for $R_{\HOI} = 10$, $R_{\HOI} = 30$ and $R_{\HOI} = 100$ are very similar, with the effect of the modification on the VMEOF getting stronger as $\omega$ increases. For negative values of $\beta$, the VMEOF decreases, whereas for positive values of $\beta$, the VMEOF increases. For lower values of $\omega$ ($\omega \leq 0.001$) the modification becomes more stable over time, and the dynamics align increasingly with those without an HOI, but with different interaction strength. This once again shows that the changes in oscillatory dynamics are caused by the HOI, and not by the change in strength of the pairwise interactions. In both Figs.~\ref{fig: beta MCSSA} and \ref{fig: omega MCSSA}, we see that, for high values of $R_{\HOI}$, the VMEOF for $\omega > 0.01$ does not change much as the curves for $\omega = 0.01$, $0.1$ and $1$ become almost indiscernible for the higher values of $R_{\HOI}$. This is because both~$\omega$ and $R_{\HOI}$ have a dispersing effect on the modification value $m$ over the grid, but for high values of $R_{\HOI}$, the modification is already very dispersed, so increasing $\omega$ does not have much more effect. However, for very small values of $\omega$ (\emph{e.g.,} $\omega = 0.001$), the modification $m$ does not only become more homogeneous in the spatial scale, but also temporally. 

For $R_{\HOI} = 3$, however, the results are different. For $\beta < 0$, the VMEOF first decreases with increasing $\omega$, but then increases again for higher values of~$\omega$. For positive values of~$\beta$, the VMEOF first increases and later decreases with increasing~$\omega$. : "We observe a maximum VMEOF for $\omega \approx 0.01$, consistent with mean-field predictions~\cite{vangielModificationSpeedAlters2025}. However, our ABM reveals that this temporal sensitivity is intrinsically constrained by the spatial scale. As $R_{\HOI}$ increases, the distinct peak flattens (Fig.\ref{fig: omega MCSSA}d), demonstrating that spatial dispersion can mitigate the oscillatory effects of temporal delays—a stabilizing mechanism that purely temporal models cannot capture. 

\section{Conclusion}
In this paper, we have shown that an HOI can have a strong effect on the dynamics of a three-species system with intransitive competition. We have shown that the strength of the HOI can impact the mean population abundances of the species, and the oscillatory dynamics. The model was designed in such a way so the speed and radius of the HOI do not have any effect on the mean population abundances. However, even though they don't affect the mean population abundances, they have a significant impact on the oscillatory dynamics of the system. In general, for a modification that lowers the mean probability of competition, the amplitude of the oscillations decreases, whereas for a modification that increases the probability of competition, the amplitude of oscillations in the system increases. These effects are most pronounced for systems with medium to high modification speeds~$\omega$, depending on the modification radius~$R_{\HOI}$. For higher values of $R_{\HOI}$, $\omega$ has a less pronounced effect if it is not very small. For very small values of~$\omega$, the system behaves rather like a system without an HOI, and the change in oscillatory dynamics due to the interaction modification is much less pronounced. Overall, these results show that the dynamics introduced by the HOI can have a strong effect on the system which would not be there with pairwise interactions only.

\appendix
\section{Methods}
\subsection*{Simulation details}  
A full ODD protocol~\cite{grimmODDProtocolDescribing2020} can be found in the supplementary material, Sec.~S1. 
Simulations were carried out using Julia~1.10.5. The code is available on GitHub: \url{https://github.com/Vahieltje/HigherOrderInteraction_ABM_paper_code} and Zenodo: \url{https://doi.org/10.5281/zenodo.17724416}. Data/project management in Julia was done with DrWatson~\cite{datserisDrWatsonPerfectSidekick2020}. The Sobol analysis was performed using GlobalSensitivity.jl~\cite{dixit2022globalsensitivity}, Plots were made using Makie~\cite{DanischKrumbiegel2021} and Inkscape.

\subsection*{Model and simulation assumptions}

Simulations are initiated with 200 agents per species (so 600 agents in total) randomly distributed over the 200 x 200 grid. The initial value of the modifier was set to $m = 1$ for all grid cells, thus assuming no initial modification. The simulations were run for 10~000 time steps, which is long enough to ensure the system has reached its equilibrium. All simulations were run for 100 repetitions without extinctions (Sec.~S2).

\subsection*{Declaration of generative AI and AI-assisted technologies in the writing process.} During the preparation of this work, T.V.G. used several AI tools: Github-copilot for help when writing the code, Grammarly for help with grammar and spelling, perplexity AI to start the literature review. After using this tool/service, the author(s) reviewed and edited the content as needed, and take(s) full responsibility for the content of the published article.

\section{Acknowledgements} 
  We are grateful to the members of the KERMIT and BionamiX research units and others at the Department of Data Analysis and Mathematical Modelling for useful and interesting discussions, tips and insights. This work was supported by the Research Foundation Flanders under grant number 3G0G0122 and a UGent-BOF GOA project ``Assessing the biological capacity of ecosystem resilience'',  grant number BOFGOA2017000601.

  The resources and services used in this work were provided by the VSC (Flemish Supercomputer Center), funded by the Research Foundation - Flanders (FWO) and the Flemish Government.

  \subsection*{Author contributions}
  Conceptualization, All authors; Software and investigation of MC-SSA, T.V.G. and H.J., Other software and investigation: T.V.G.; Validation and Writing - original draft, T.V.G.; Funding Acquisition, supervision and writing - Review \& Editing, J.M.B. and B.D.B.; supervision and writing - Review \& Editing, A.J.D. 
  
\subsection*{Competing interests}
  There are no competing interests.

 \bibliographystyle{elsarticle-num} 
 \bibliography{Citations}

\end{document}